\documentclass[aps,prl,superscriptaddress,reprint,showpacs,longbibliography,nofootinbib,amsmath,amssymb]{revtex4-1}
\usepackage[utf8]{inputenc}
\usepackage[T1]{fontenc}

\usepackage{graphicx}
\usepackage{grffile}
\usepackage{microtype}
\usepackage[usenames,dvipsnames]{xcolor}
\usepackage[colorlinks,linkcolor=blue,citecolor=blue,urlcolor=blue]{hyperref}

\def\equationautorefname~#1\null{equation~(#1)\null}

\newcommand{\bra}[1]{\langle#1|}
\newcommand{\ket}[1]{|#1\rangle}

\begin{document}

\title{Suppressing photochemical reactions with quantized light fields}
\author{Javier Galego}
\affiliation{Departamento de F{\'\i}sica Te{\'o}rica de la
  Materia Condensada and Condensed Matter Physics Center (IFIMAC),
  Universidad Aut\'onoma de Madrid, E-28049 Madrid, Spain}
\author{Francisco~J.~Garcia-Vidal}
\email{fj.garcia@uam.es}
\affiliation{Departamento de F{\'\i}sica Te{\'o}rica de la
  Materia Condensada and Condensed Matter Physics Center (IFIMAC),
  Universidad Aut\'onoma de Madrid, E-28049 Madrid, Spain}
\affiliation{Donostia International Physics Center (DIPC), E-20018 Donostia/San Sebastian, Spain}
\author{Johannes Feist}
\email{johannes.feist@uam.es}
\affiliation{Departamento de F{\'\i}sica Te{\'o}rica de la
  Materia Condensada and Condensed Matter Physics Center (IFIMAC),
  Universidad Aut\'onoma de Madrid, E-28049 Madrid, Spain}

\begin{abstract}
\end{abstract}

\date{\today}
\maketitle

\textbf{Photoisomerization, i.e., a change of molecular structure
  after absorption of a photon, is one of the most fundamental
  photochemical processes. It can perform desirable functionality,
  e.g., as the primary photochemical event in human vision, where it
  stores electronic energy in the molecular
  structure~\cite{Yoshizawa1963,Polli2010}, or for possible
  applications in solar energy storage~\cite{Kucharski2011} and as
  memories, switches, and actuators~\cite{Irie2014,Guentner2015}; but
  it can also have detrimental effects, for example as an important
  damage pathway under solar irradiation of
  DNA~\cite{Sinha2002,Douki2003}, or as a limiting factor for the
  efficiency of organic solar cells~\cite{Zietz2014}. While
  photoisomerization can be avoided by shielding the system from
  light, this is of course not a viable pathway for approaches that
  rely on the interaction with external light (such as solar cells or
  solar energy storage). Here, we show that strong coupling of organic
  molecules to a confined light mode can be used to strongly suppress
  photoisomerization, and thus convert molecules that normally show
  fast photodegradation into photostable forms.}

Strong coupling is achieved when the coherent energy exchange between
molecules and the light mode becomes faster than the decoherence
processes in the system~\cite{Kaluzny1983,Thompson1992}. This creates
paradigmatic hybrid quantum systems with eigenstates that have mixed
light-matter character (so-called polaritons). Organic materials
provide particularly large dipole moments and high molecular
densities, making them ideal systems to reach the strong coupling
regime~\cite{Lidzey1998,Bellessa2004}. By exploiting the strong field
localization in plasmonic nanocavities, even single-molecule strong
coupling has recently been achieved~\cite{Chikkaraddy2016}. However,
while most models of strong coupling are based on two-level systems,
this is far from a realistic description for molecules with many
nuclear (i.e., rovibrational) degrees of freedom. While pioneering
experiments show modifications of material properties and even
chemical reaction rates under strong coupling
\cite{Hutchison2012,Wang2014a,Coles2014,Orgiu2015}, the influence of
strong coupling on internal degrees of freedom has only come into
focus recently~\cite{Spano2015,Galego2015,Cwik2016,
  Herrera2016,Kowalewski2016}. We here demonstrate that a wide class
of photochemical reactions can be strongly suppressed under strong
coupling to discrete quantized light modes. Our results imply that
even extremely fragile molecules could be stabilized by simply putting
them close to a nanophotonic structure.

\begin{figure}
  \includegraphics[width=\linewidth]{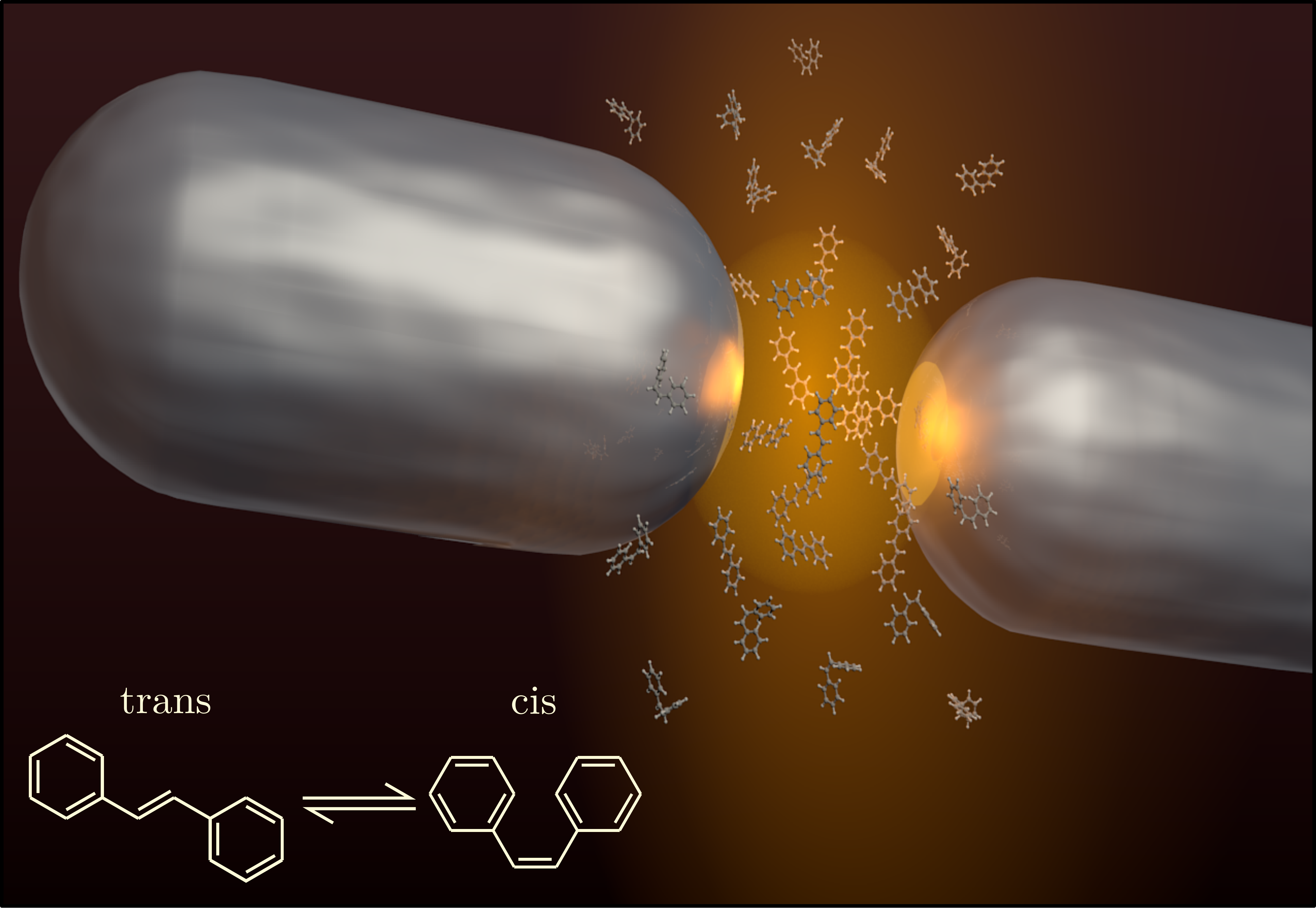}
  \caption{Schematic of a collection of molecules coupled to a
    localized surface plasmon mode in the gap between two
    nanoparticles.}
  \label{fig:setup}
\end{figure}

\begin{figure*}
  \includegraphics[width=\linewidth]{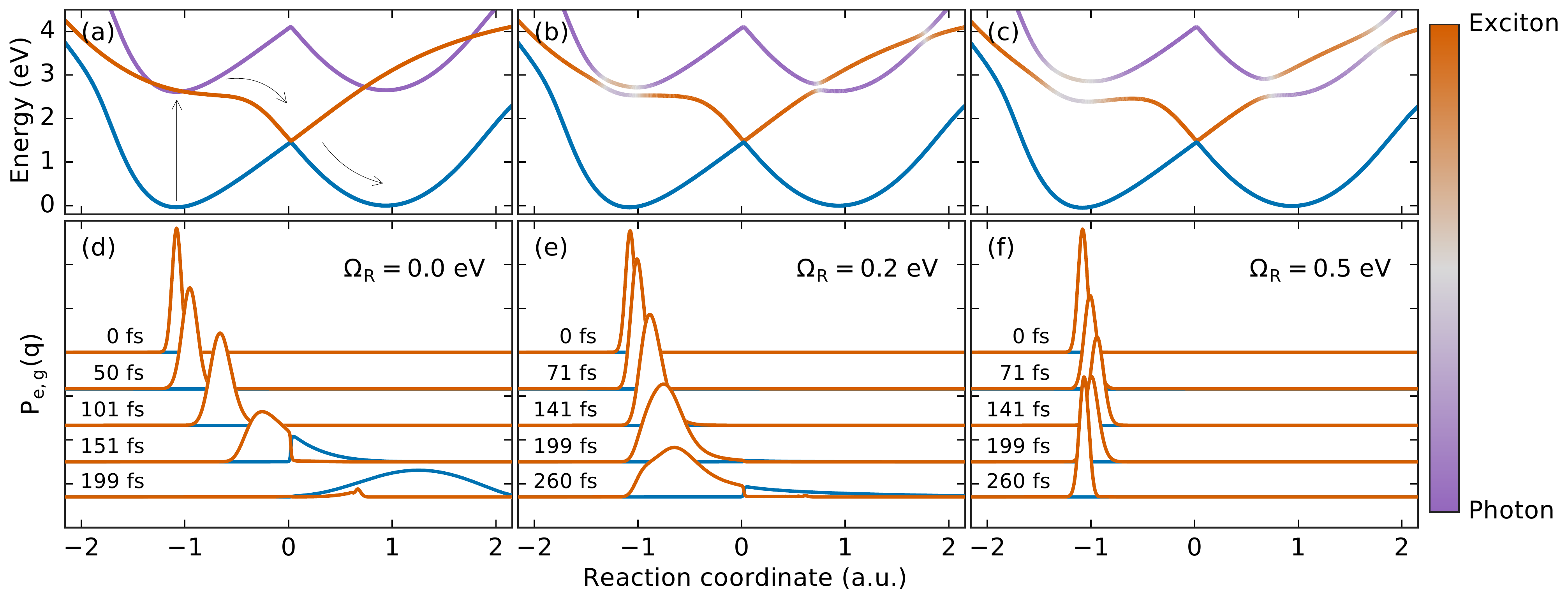}
  \caption{Suppression of photoisomerization under strong coupling for
    a single molecule. (a-c) Ground (blue) and excited (purple-orange
    color scale) potential energy surfaces of the model molecule
    coupled to a quantized light mode ($\omega_c=2.65$~eV), with the
    light-matter coupling strength $\Omega_R$ increasing from (a) to
    (c). The continuous color scale encodes the nature of the
    hybridized excited PES. (d-e) Time propagation of the nuclear
    wavepacket after sudden excitation to the lowest excited PES
    (=lower polariton for $\Omega_R>0$), shown separately for the
    parts in the ground and lower polariton surface (orange), and the
    ground state surface (blue) reached through the nonadiabatic
    transition at $q=0$. Contributions in the upper polariton surface
    are negligible and not shown.}
  \label{fig:single_molecule_timeprop}
\end{figure*}

We treat a general molecular model that can represent a variety of
commonly studied photoisomerization reactions, such as
\emph{cis-trans} isomerization of stilbene, azobenze or
rhodopsin~\cite{Waldeck1991,Polli2010,Quick2014} (corresponding to
rotation around a C=C or N=N double bond, as sketched in the inset of
\autoref{fig:setup}), or ring-opening and ring-closing reactions in
diarylethenes~\cite{Irie2014}. The model molecule (see Methods section
for more detail) describes nuclear motion on ground and excited
electronic potential energy surfaces (PES) along a single reaction
coordinate $q$, as shown in
\autoref{fig:single_molecule_timeprop}a. All other degrees of freedom
are assumed to be fully relaxed, such that the excited PES represents
the minimum-energy reaction path. The ground state PES possesses two
minima separated by a barrier, corresponding to the two isomers. At
the top of the barrier, a narrow avoided crossing between the ground
and excited PES leads to an efficient nonadiabatic transition, giving
a photoisomerization quantum yield approaching unity. As shown in
\autoref{fig:single_molecule_timeprop}d, the bare model molecule
undergoes rapid photoisomerization within a few hundred fs. We here
perform full wavepacket propagation after excitation from the ground
to the excited electronic state by an ultrashort laser pulse.

In contrast, when the system enters strong coupling,
photoisomerization in a single molecule is suppressed. To show this,
we rely on the theoretical framework we introduced in
ref.~\cite{Galego2015}, which extends the well-known Born-Oppenheimer
approximation with the tools of cavity QED by including the
light-matter interaction in the ``electronic'' Hamiltonian and
following nuclear dynamics on hybrid light-matter PES. We include a
(single) quantized light mode (which can represent confined light
modes in different physical systems, such as microcavity modes or
localized surface plasmon resonances) with energy term
$\omega_c \hat{a}^\dagger \hat{a}$. Here, $\omega_c$ is the quantized
mode frequency, and $\hat{a}^\dagger$ and $\hat{a}$ are the associated
bosonic creation and annihilation operators. The light-matter coupling
is given by
$\hat\mu(q) \cdot \vec E_{1\mathrm{ph}} (\hat{a}^{\dagger}+\hat{a})$,
where $\vec E_{1\mathrm{ph}}$ is the electric field amplitude of a
single quantized confined photon, and $\hat{\mu}$ is the (vectorial)
molecular dipole operator. Without light-matter coupling, the
photonically excited surface $V_c(q) = V_g(q)+\omega_c$ is simply a
copy of the molecular ground state shifted upwards by the photon
energy (\autoref{fig:single_molecule_timeprop}a). When coupling is
turned on, the two singly excited surfaces $V_c(q)$ and $V_e(q)$
hybridize, forming ``polaritonic'' surfaces with mixed light-matter
character, as depicted in
\autoref{fig:single_molecule_timeprop}(b,c). The splitting between the
polaritonic PES around equilibrium ($q_0\approx-1.05~$a.u.) is
approximately equal to the Rabi frequency
$\Omega_R=2\vec\mu_{eg}(q_0)\cdot\vec E_{1\mathrm{ph}}$. Importantly,
the lower polariton PES develops a deeper and deeper minimum as the
coupling is increased. This has two primary reasons: First, the
light-matter coupling is most effective when $V_c(q)$ and $V_e(q)$ are
close, ``pushing down'' the lower polariton. At regions of larger
detuning, the ``polariton'' PES are almost identical to the uncoupled
ones. Second, the local shape of the polariton PES becomes a mixture
of the two uncoupled PES in regions where they hybridize
significantly. Since the photonic surface $V_c(q)$ behaves like the
ground-state PES, this additionally supports the formation of a local
minimum in the polaritonic PES. In combination, this provides a
reaction barrier that almost completely suppresses isomerization for
sufficiently strong coupling, as seen in
\autoref{fig:single_molecule_timeprop}(b,c). Note that while the upper
polariton PES appears even more stable than the lower one in this
model, this is an artefact of the restriction to one degree of
freedom, with all other degrees of freedom relaxed to their local
minimum. This implies that the lower polariton PES indeed corresponds
to the lowest-energy excited state, such that the restriction to one
coordinate is well-justified. In contrast, the upper polariton surface
can possess efficient relaxation pathways to the lower polariton along
orthogonal degrees of freedom, and indeed, upper polaritons are known
to decay relatively quickly within the excited-state
subspace~\cite{Litinskaya2004,Coles2011}.

We have thus shown that strong coupling of a single molecule to a
confined light mode can strongly suppress photoisomerization reactions
and stabilize the molecule. The recent experimental realization of
single-molecule strong coupling proves that this could indeed be a
viable pathway towards manipulation of single
molecules~\cite{Chikkaraddy2016}. At the same time, most experiments
achieving strong coupling with organic molecules have exploited
\emph{collective} coupling~\cite{Wiederrecht2004,Zengin2015}, in which
$N\gg1$ molecules coherently interact with a single mode, leading to
an enhancement of the total Rabi frequency by a factor of
$\sqrt{N}$~\cite{Torma2015}. However, it has recently been shown that
many observables corresponding to ``internal'' degrees of freedom of
the molecules are only affected by the single-molecule coupling
strength~\cite{Galego2015,Cwik2016}. One could thus expect that the
suppression of photoisomerization disappears under collective strong
coupling when $N$ is sufficiently large. We next show that exactly the
opposite is the case, and strong coupling of a large number of
molecules to a single mode actually improves the molecular
stabilization significantly.

\begin{figure*}
  \includegraphics[width=\linewidth]{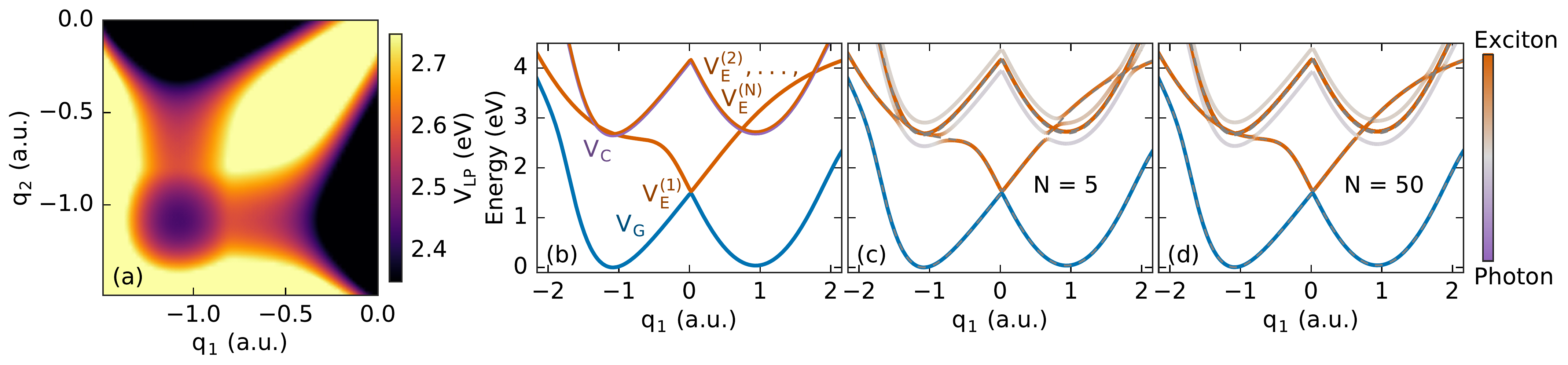}
  \caption{Many-molecule potential energy surfaces under strong
    coupling. (a) Lower polariton PES for $N=50$ molecules, under
    motion of molecules $1$ and $2$, with all others held in the
    equilibrium position $q_0$. (b-d) All potential energy surfaces
    under motion of only molecule $1$, for no light-matter coupling
    (b), and under strong coupling for $N=5$ (c) and $N=50$ (d)
    molecules. In all panels, the photonic mode frequency is
    $\omega_c=2.65$~eV, while the (collective) Rabi frequency is fixed to
    $\Omega_C = \sqrt{N} \Omega_R = 0.5~$eV.}
  \label{fig:many_molecule_pes}
\end{figure*}

In order to treat collective strong coupling involving $N$ molecules
and a single confined light mode, we again restrict ourselves to the
zero- and single-excitation subspace. The molecules now have $N$ total
nuclear degrees of freedom, described by the vector
$\vec q = (q_1,\ldots,q_N)$, and the PES accordingly become
$N$-dimensional surfaces. For the uncoupled system, these surfaces are
the global ground state $V_G(\vec q) = \sum_i V_g(q_i)$, the
photonically excited state $V_C(\vec q)=V_G(\vec q)+\omega_c$, and the
$N$ molecular excited states
$V_E^{(i)}(\vec q) = V_G(\vec q) + V_e(q_i)-V_g(q_i)$. The
electronic-photonic Hamiltonian in the first excited subspace is then
given by
\begin{equation} \label{eq:HBO_many}
  \hat{H}(\vec q) = \begin{pmatrix}
    V_C(\vec q) & g(q_1) & g(q_2) & \ldots & g(q_N)\\[2pt]
    g(q_1) & V_E^{(1)}(\vec q) & 0 & \ldots & 0\\[2pt]
    g(q_2) & 0 & V_E^{(2)}(\vec q) & \ldots & 0\\[2pt]
    \vdots & \vdots & \vdots & \ddots & \vdots\\[2pt]
    g(q_N) & 0 & 0 & \ldots & V_E^{(N)}(\vec q)\\[2pt]
  \end{pmatrix},
\end{equation}
where $g(q) = \vec\mu_{eg}(q)\cdot\vec E_{1\mathrm{ph}}$.
Diagonalizing $\hat{H}(\vec q)$ gives $N+1$ polaritonic surfaces,
which describe the collective coupled motion of all molecules. In
principle, this could induce, e.g., collective transitions in which
multiple molecules move in concert. We show in
\autoref{fig:many_molecule_pes}a that this is not the case. Here, we
plot the lower-polariton PES (the lowest excited-state surface) as
a function of the reaction coordinates of the first two molecules,
$q_1$ and $q_2$, while keeping all other molecules fixed to the
equilibrium position ($q_j=q_0$ for $j>2$). The figure clearly shows
that the barrier for isomerization starting from the ground-state
equilibrium position $\vec q = (q_0,\ldots,q_0)$ is minimal for motion
along only one molecular degree of freedom.

We thus analyze the polaritonic states under motion of only the first
molecule $q_1$, fixing all other molecules to the ground-state
equilibrium position ($q_j=q_0$ for $j>1$). The corresponding PES are
shown in \autoref{fig:many_molecule_pes}(b-d). When the light-matter
coupling is zero (\autoref{fig:many_molecule_pes}b), the surface
$V_E^{(1)}(\vec q)$ behaves like $V_e(q_1)$, while all other surfaces
(corresponding to photonic excitation, or excitation of a
``stationary'' molecule $j>1$) appear like copies of the ground-state
PES $V_g(q_1)$ shifted in energy.

The strongly coupled PES for varying numbers of molecules are shown in
\autoref{fig:many_molecule_pes}(c,d). We keep the total Rabi frequency
constant (corresponding to a scaling of the single-photon field
strength with $N^{-1/2}$). Close to equilibrium ($q_1\approx q_0$),
the $N+1$ surfaces can be clearly classified into a lower and upper
polariton PES, which show significant hybridization with the photonic
mode, as well as $N-1$ ``dark'' surfaces that are almost purely
excitonic~\cite{Galego2015}.

As the number of molecules is increased, the local minimum of the
lower-polariton PES becomes more and more reminiscent of the pure
ground-state PES, making the potential energy barrier to
photoisomerization higher and higher. This can be immediately
understood from the structure of the Hamiltonian: For motion along any
given molecular degree of freedom, there is only one PES that supports
photoisomerization, but $N$ PES that give ground-state-like motion
($N-1$ molecular and one photonic excitation). The lower polariton PES
inherits its shape from these ingredients, weighted by their
respective fractions. This results in an almost perfect copy of the
ground state PES for motion along any one molecular degree of freedom,
stabilizing the molecules through collective protection of the
excitation by distributing it over all molecules.

Furthermore, the similarity of the ground and lower polariton PES for
large $N$ implies that the Franck-Condon factors, i.e., the overlap
between nuclear eigenstates in the ground and lower polariton PES,
become approximately diagonal. Thus, transitions from the overall
ground state to vibrationally excited states in the lower polariton
PES become more and more suppressed, such that the excited nuclear
wavepacket is in the ground state, providing an additional
stabilization effect.

Finally, a third effect further improves the stabilization in the
lower polariton. Closer inspection of
\autoref{fig:many_molecule_pes}(c,d) reveals that the lower polariton
PES features a narrow avoided crossing (at $q\approx-0.75$) where it
switches from a hybridized state to essentially the single-molecule
excited-state surface. The large wavefunction mismatch makes adabiatic
nuclear motion unlikely, and diabatic motion, in which the
electronic and photonic degrees of freedom are unchanged, becomes much
more likely. This can be shown by constructing diabatic PES close to
the avoided crossing, obtained by diagonalizing the coupling between
$N-1$ ``unmoving'' molecules and the light mode (giving a very good
approximation to the LP PES), which is then coupled to the
excited-state PES of the single moving molecule. A short calculation
reveals that the transition matrix element between the diabatic LP
surface and the single-molecule excited surface is surpressed by a
factor $\sim N^{-1/2}$ for a fixed collective Rabi frequency,
indicating that the transition to the isomerization surface is
indeed strongly suppressed.

To conclude, we have demonstrated the stabilization of excited-state
molecular structure and accompanying strong suppression of
photochemical reactions under strong coupling of molecules to confined
light modes. While already effective in the case of a single couple
molecule, we find that collective coupling of a large number molecules
to a single light mode does not reduce the influence of strong
light-matter coupling on each molecule, but provides even stronger
stabilization. This counterintuitive feature can be understood by the
additional protection afforded by collective distribution of the
excitation over the molecules. These results do not depend on the
specifics of the molecular model, such that the observed stabilization
is expected to occur for any kind of photochemical reaction that is
induced by motion on the excited molecular PES. These results thus
pave the way towards a new type of material, created through strong
coupling to quantized light modes, for devices such as solar cells.

\section{Methods}
We here describe the molecular model in more detail. The adiabatic PES
of the bare molecule are constructed in terms of diabatic surfaces
$V_A(q)$ and $V_B(q)$ coupled to each other with a coupling $h_0$ that
is assumed constant in space. This gives the following electronic
Hamiltonian:
\begin{equation} \label{eq:Helec_diab}
\hat{H}_{\mathrm{el}}(q) = \begin{pmatrix} V_A(q) & h_0 \\ h_0 & V_B(q) \end{pmatrix}.
\end{equation}
Diagonalization of $\hat{H}_{\mathrm{el}}(q)$ returns the ground and
excited state PES of \autoref{fig:single_molecule_timeprop}a, $V_g(q)$
and $V_e(q)$, together with the adiabatic electronic
wavefunctions. This also gives access to the nonadiabatic coupling
that controls the transition between ground and excited surfaces at
$q \approx 0$, given by $F_{i,j}(q) = \bra{i(q)}\partial_q\ket{j(q)}$,
where $i,j = e,g$ and $\ket{i(q)}$ represent the adiabatic electronic
states.  We note that nonadiabatic transitions in ``real'' molecules
typically involve conical intersections~\cite{Worth2004}, which only
occur in multi-dimensional systems; however, the details of this
transition do not influence the results presented.

The complete molecular Hamiltonian is then given by
\begin{equation} \label{eq:Hmol}
\hat{H}_{\mathrm{mol}}(q) =  \frac{\hat{P}^2}{2M_q} + \hat{V}(q) + \hat{\Lambda}(q),
\end{equation}
where $\hat{P}$ is the (diagonal) nuclear momentum operator, $M_q$ is
the effective mass for the nuclear coordinate $q$, $\hat{V}(q)$ is
the (diagonal) PES matrix in the adiabatic basis, and
$\hat{\Lambda}(q)$ is the matrix of off-diagonal (nonadiabatic)
couplings, given by
$\hat{\Lambda}(q) = \frac{1}{2M_q}\left (2\hat{F}(q) \partial_q +
  \hat{G}(q) \right)$,
with $G_{i,j}(q) = \partial_q F_{i,j}(q) + F_{i,j}^2(q)$ \cite{Worth2004}.

When introducing the coupling to the quantized confined light mode,
the total Hamiltonian additionally depends on the dipole moment
$\hat{\mu}(q)$, which we set as purely offdiagonal in the adiabatic
basis. The ground-excited dipole moment $\vec\mu_{eg}(q)$ typically is
approximately constant close to the stable geometries, but changes
rapidly close to the nonadiabatic transition due to the sudden
polarization effect \cite{Bonacic-Koutecky1975}. We thus choose
$\mu_{eq}(q)\propto \arctan(q/q_m)$, with $q_m=0.625$ representing the
length scale on which $\mu_{eq}(q)$ changes. Diagonalization of the
total adiabatic $N$-molecule electron-photon Hamiltonian
\begin{equation}
  \hat{H}_{\mathrm{SC}} = \omega_c a^\dagger a +
  \sum_i \left(\hat{V}(q_i) + \hat\mu(q_i) \cdot \vec E_{1\mathrm{ph}}
    (\hat{a}^{\dagger}+\hat{a}) \right)
\end{equation}
within the single-excitation subspace then yields the strongly coupled
(polaritonic) potential energy surfaces. We note that the nonadiabatic
couplings in the polaritonic basis are given by new terms
$\hat{\Lambda}_{SC}$ due to the basis change, as well as the
bare-molecule nonadiabatic couplings $\hat{\Lambda}(q_i)$ transformed
to the polaritonic basis.

To evaluate population transfer both in the uncoupled and in the
strongly coupled system, we finally solve the time-dependent
Schrödinger equation
$i\partial_t \ket{\psi(t)} = \hat{H}_{\mathrm{tot}} \ket{\psi(t)}$
without invoking the Born-Oppenheimer approximation, i.e., including
all nonadiabatic terms. The initial wavefunction is given by direct
promotion of the ground-state nuclear wavepacket to the lowest excited
state (excited molecular state for no coupling, lower polariton under
strong coupling), filtered by the $q$-dependent transition dipole
moment from the ground state. This is the initial state that would be
obtained after excitation by an ultrashort laser pulse tuned to the
excitation energy around the nuclear equilibrium position.

\section{Acknowledgements}
This work has been funded by the European Research Council
(ERC-2011-AdG proposal No. 290981), by the European Union Seventh
Framework Programme under grant agreement FP7-PEOPLE-2013-CIG-618229,
and the Spanish MINECO under contract MAT2014-53432-C5-5-R and the
``María de Maeztu'' programme for Units of Excellence in R\&D
(MDM-2014-0377).

\bibliography{references}

\end{document}